\documentclass[twocolumn,prd,showpacs,superscriptaddress,groupedaddress,preprintnumbers,nofootinbib]{revtex4-1}
\pdfoutput=1
\usepackage{graphicx,amsfonts,amsmath,amssymb,epsfig,color, mathrsfs,latexsym,url,wasysym,float}
\newcommand{\be}{\begin{equation}}
\newcommand{\ee}{\end{equation}}
\newcommand{\nn}{\nonumber}
\newcommand{\ba}{\begin{eqnarray}}
\newcommand{\ea}{\end{eqnarray}}
\newcommand{\mpl}{m_{\rm Pl}}
\begin{document}

\title{Emergent Weak Scale from Cosmological  Evolution and Dimensional Transmutation}

\author{Ahmad Sadeghi}\author{Mahdi Torabian$^*$}
\affiliation{Department of Physics, Sharif University of Technology, Azadi Ave, 11155-9161, Tehran, Iran} 
\affiliation{School of Particles and Accelerators, Institute for Research in Fundamental Sciences (IPM), 19395-5531, Tehran, Iran}

\begin{abstract}
In this note we present a framework in which the weak scale appears dynamically technically natural with no new physics up to the Planck scale. The mixing between the massless Higgs and the $R^2$ metric theory induces, in canonical parametrization of the Einstein frame, an effective field-dependent Higgs mass parameter. It is a dynamical variable which in the course of cosmic evolution scans a wide range of values and eventually stabilizes at a low scale. The one-loop effective potential has an electroweak symmetry breaking vacuum and the hierarchy is explained by dimensional transmutation and cosmological relaxation mechanisms. Furthermore, by evaluating the renormalization group improved effective potential we find that the electroweak vacuum is the global minimum of the effective potential.
\end{abstract}

\pacs{14.80.Bn, 12.60.Fr, 98.80.Cq}
\maketitle

\subsection*{Introduction} 
The recent combined ATLAS and CMS analyses of the Higgs properties yield a mass slightly above 125 GeV and other qualities (couplings, $J^{\tt CP}$, etc) in excellent agreement with the Standard Model (SM) predictions \cite{Aad:2015zhl}. 
The discovery of the Higgs boson \cite{Aad:2012tfa,Chatrchyan:2012ufa}, a light elementary fermiophilic scalar particle, has completed the SM and at the same time indicates its incompleteness and calls for beyond the SM. The SM could in principle be employed to make predictions at an arbitrary energy scale. However, the UV sensitive large difference between the pole mass and the renormalized ${\overline{\rm MS}}$ mass demands fine-tuning of parameters to explain the hierarchy between the electroweak (EW) scale, set explicitly by the mass parameter, against any higher mass scale. Besides, the absolute stability of the EW vacuum is excluded at $98\%$ C.L. \cite{Degrassi:2012ry} which particularly would be problematic for large Higgs values (in the early Universe). Agnostically speaking, experiments thus far have merely explored the shape of the Higgs potential very close the origin and the mechanism for the EW symmetry breaking (EWSB) and the global shape of the scalar potential in not yet known. The latter is of particular importance when one studies the Higgs dynamics in relation to the physics of the early Universe.

Motivated by the above lines, in this note, we study the non-minimal coupling of a massless Higgs field to an $R^2$ gravity. Cosmological data indicate that this theory of gravity could be relevant to the early times;  {\it i.e.} inflationary observables are consistent with the predictions of the $R^2$ theory \cite{Ade:2015xua}. In the Jordan frame parametrization no explicit bare or renormalized Higgs mass parameter is introduced in the action. Nevertheless, in the canonical Einstein frame parametrization (where a new propagating scalar field appears though the Weyl transformation) the Higgs receives an effective Weyl-field dependent mass-squared parameter. The mass parameter is then a dynamical variable and scans over a vast range of values as the Weyl field evolves in the course of cosmic evolution. Although the tree-level potential has minima at zero, we find that the one-loop effective potential stabilizes both fields at non-zero values. The EW symmetry is radiatively spontaneously broken and a new scale, parametrically smaller than the Planck scale, is produced via dimensional transmutation a la Coleman-Weinberg mechanism \cite{Coleman:1973jx} assisted by gravitational couplings. The weak scale is then dynamically technically natural and the hierarchy is explained by quantum renormalization group (RG) running and classical Weyl field excursion. Furthermore, we compute the one-loop RG improved effective potential and we find that the EW vacuum is the global minimum of the potential. 

Recently, the idea of cosmological relaxation has received great interests where the Higgs-axion interplay has been applied to explain the hierarchical scales \cite{Graham:2015cka,Espinosa:2015eda,Gupta:2015uea,Kaplan:2015fuy,Hardy:2015laa,Jaeckel:2015txa,Matsedonskyi:2015xta,Marzola:2015dia,Ibanez:2015fcv}. In fact, the hierarchy is explained by a technically natural symmetry-breaking small parameter. In the present work, the Higgs-gravity interplay has been examined to explain the large hierarchy between the weak scale and the Planck scale. This framework is minimal in the sense that a cosmic inflationary era is guaranteed and the pre/post inflationary dynamics of the Higgs field is connected to the present EW vacuum.

\subsection*{The Higgs/Gravity Reciprocity}  
We study non-minimal coupling of the Higgs field to the $R^2$ theory of gravity whose dynamics in the Jordan frame is given by the following action
\ba S=\int {\rm d}^4x(-g)^{1/2}\frac{1}{2}\Big(&&(M_J^2+\xi\varphi^2) R + \alpha R^2 \cr &&- g^{\mu\nu}\partial_\mu\varphi\partial_\nu\varphi - 2V_J(\varphi)\Big),\ea
where $\varphi^2=2H^\dagger H$ in the unitary gauge and $V_J$ is the Higgs potential. The non-minimal coupling parameter $\xi$ can have either sign, with the above convention it is assumed positive. We can make the kinetic term of the metric canonical through a local Weyl transformation as 
\be g_{\mu\nu}^E = (M_J^2+\xi\varphi^2 + 2\alpha R)\mpl^{-2} g_{\mu\nu} \equiv e^{\tilde\chi} g_{\mu\nu},\ee
which gives the action in the so called Einstein frame 
\ba\label{Einstein-frame-action} S_E = \int {\rm d}^4x(-g_E)^{1/2}\Big(&&\frac{1}{2} \mpl^2R_E - \frac{1}{2}g_E^{\mu\nu}\partial_\mu\chi\partial_\nu\chi \cr - &&\frac{1}{2}e^{-\tilde\chi}g_E^{\mu\nu}\partial_\mu\varphi\partial_\nu\varphi  -V_E(\varphi,\chi)\Big)\!.\ \ \ea 
There appears a new propagating scalar field called the Weyl field $\chi=\tilde\chi\mpl\sqrt{3/2}$. Furthermore, the scalar potential in the this new parametrization reads as follows
\ba\label{potential} V_E(\varphi,\chi)\!\!=\! e^{-2\tilde\chi}\Big(V_J(\varphi)+\frac{1}{8}\alpha^{-1}\mpl^4\big(e^{\tilde\chi}-1-\!\xi\mpl^{-2}\varphi^2\big)^2\Big).\nn\\\ea
We assume no Higgs mass parameter in the potential and only quartic self-interaction with coupling $\lambda_J$. Then, the tree-level scalar potential in the Einstein frame takes the following form
\ba\label{potential} V_E(\varphi,\chi)= \frac{1}{8}\alpha^{-1}\big(1-e^{-\tilde\chi}\big)^2\mpl^4 + \frac{1}{2}\mu_{\rm }^2\varphi^2 + \frac{1}{4}\lambda\varphi^4,\ \quad\ea
where we have applied a compact notation
\ba\label{eff-lambda} \lambda &\equiv& e^{-2\tilde\chi}(\lambda_J+\frac{1}{2}\xi^2\alpha^{-1}), \\\label{eff-mass} \mu_{\rm }^2&\equiv&-\frac{1}{2}e^{-2\tilde\chi}\xi\alpha^{-1}(e^{\tilde\chi}-1)\mpl^2.\ea
The first term in the above potential is Higgs independent and flat for large Weyl field values and thus drives cosmic inflation \cite{Starobinsky:1980te}. The cosmological observations fix the parameter $\alpha$ to about $1.5\times 10^9$ \cite{Ade:2015xua}.

It can be seen from above that the gravitational couplings, besides altering the Higgs quartic coupling, induce an effective Hiss mass-squared parameter $\mu^2$. It is a field-dependent dynamical variable and changes as the Weyl field varies. 
Indeed, in the course of cosmic history while the Weyl field relaxes to its minimum, it scans a wide range of values from about the Planck scale to zero. This parameter is negative for our positive choice of $\xi$ and can get the Higgs a non-zero vacuum expectation value (VEV). For the same reason the Higgs VEV also scans a broad range parametrically from $\varphi_0^2 \sim \mpl^2\xi^{-1}(e^{\tilde\chi}-1)$ to zero. If the Weyl field stabilizes at a non-zero value, then there will be a local EWSB minimum. It can be easily seen that the minimum of the tree-level potential in both Higgs and Weyl directions is at zero; neither of them receive a non-zero VEV. In order to thoroughly look at the vacuum structure and possible EWSB vacua in the next section we compute the one-loop effective potential.

\subsection*{One-Loop Effective Potential}
With the notation adopted above, the one-loop correction to the tree-level potential \eqref{potential} can be easily computed to give the familiar Coleman-Weinberg effective potential in the Landau-'t Hooft gauge
\be\label{effpot} V_{E}^{\rm eff}=\frac{1}{2}\mu_R^2\varphi^2 \!+\!\frac{1}{4}\lambda_R\varphi^4 \!+\!\frac{1}{64\pi^2}\!\!\sum_i d_i m_i^4\big[\ln\!\big(m_i^2/M^2\big)\!-\!c_i\big],\ee
where $m_i$ is the tree-level mass expression for fields couple to the Higgs, $c_i, d_i$ are fixed by given parcel species. In the above $M$ is the renormalization scale and $\lambda_R, \mu_R$ are the renormalized values of the corresponding parameters at $M$ (having understood that, we drop the index $R$ from this point on) \cite{Coleman:1973jx,Sher:1988mj}.

\subsubsection*{Electroweak Vacuum Symmetry Breaking Vacua}
For the sake of simplicity, we rewrite the effective potential \eqref{effpot} in the small field approximation as follows\footnote{Explicitly, and the one-loop corrected mass parameter and
the quartic coupling are computed as
\be\begin{split} &\lambda_{\rm eff} = \lambda + \frac{1}{16\pi^2}3 \Big[\frac{1}{16}\big(2g^4+(g^2+g'^2)^2\big)-y_t^4\Big] \ln\Big(\frac{\varphi^2}{M^2}\Big) \cr  &\qquad\qquad+ \frac{1}{16\pi^2}3\lambda^2\Big[3\ln\Big(\frac{3\lambda\varphi^2+\mu^2}{M^2}\Big) +\ln\Big(\frac{\lambda\varphi^2+\mu^2}{M^2}\Big)\Big].\cr
&\mu_{\rm eff}^2 = \mu^2 \Big(1+ \frac{1}{16\pi^2}3\lambda\Big[\!\ln\Big(\frac{3\lambda\varphi^2+\mu^2}{M^2}\Big)\!+\ln\Big(\frac{\lambda\varphi^2+\mu^2}{M^2}\Big)\Big] \Big),\end{split}\nn\ee}
\be V_{E}^{\rm eff}= \frac{1}{2}\mu_{\rm eff}^2(\varphi,\chi)\varphi^2 +\frac{1}{4}\lambda_{\rm eff}(\varphi,\chi)\varphi^4,\ee
The minima of the effective potential can be found by looking at solutions to $\partial_\varphi V_{\rm E}^{\rm eff}=0$ and $\partial_\chi V_{\rm E}^{\rm eff}=0$ which respectively give the following equations
\ba\label{ext-1} 0&=& \mu_{\rm eff}^2+\frac{1}{2}\frac{\partial\mu_{\rm eff}^2}{\partial\ln\varphi}+\Big(\lambda_{\rm eff}+\frac{1}{4}\frac{\partial\lambda_{\rm eff}}{\partial\ln\varphi}\Big)\varphi^2,\\ \label{chi-extremum}
0 &=& -\frac{1}{\sqrt{6}\xi}\mu^2\mpl + \frac{1}{2}\frac{\partial\mu_{\rm eff}^2}{\partial\chi}\varphi^2 + \frac{1}{4}\frac{\partial\lambda_{\rm eff}}{\partial\chi}\varphi^4.\ea

Furthermore, the elements of the mass matrix about the extrema are computed as follows
\ba V_{\varphi\varphi} &= &-2\mu^2 + \frac{\partial\lambda_{\rm eff}}{\partial\ln\varphi}\varphi^2 \\ &+&\frac{1}{2}\Big(\frac{\partial\mu_{\rm eff}^2}{\partial\ln\varphi}+\frac{\partial^2\mu_{\rm eff}^2}{\partial\ln\varphi^2}\Big) + \frac{1}{4}\Big(\frac{\partial\lambda_{\rm eff}}{\partial\ln\varphi}+\frac{\partial^2\lambda_{\rm eff}}{\partial\ln\varphi^2}\Big)\varphi^2,\cr
V_{\chi\chi} &=& \frac{1}{6\alpha}(2e^{-2\tilde\chi}-e^{-\tilde\chi})+\frac{1}{2}\frac{\partial^2\mu_{\rm eff}^2}{\partial\chi^2}\varphi^2+\frac{1}{4}\frac{\partial^2\lambda_{\rm eff}}{\partial\chi^2}\varphi^4,\\
V_{\varphi\chi} &=& \frac{\partial\mu_{\rm eff}^2}{\partial\chi}\varphi + \frac{\partial\lambda_{\rm eff}}{\partial\chi}\varphi^3 +\frac{1}{2}\frac{\partial^2\mu_{\rm eff}^2}{\partial\varphi\partial\chi}\varphi^2+\frac{1}{4}\frac{\partial^2\lambda_{\rm eff}}{\partial\varphi\partial\chi}\varphi^4.\qquad\ea
They can be easily diagonalized to give the mass eigenvalues. The mass of the heavy eigenstate is
\be m_\chi^2 = \frac{1}{6\alpha}\mpl^2 + {\cal O}(v\mpl), \ee
and the mass of the light eigenmode, the physical Higgs, is read as follows
\be m^2_h = V_{\varphi\varphi} - V^2_{\varphi\chi}V^{-1}_{\chi\chi} + {\cal O}(v^4/\mpl^2)
.\ee

\begin{center}\begin{figure}[b!]\includegraphics[scale=.245]{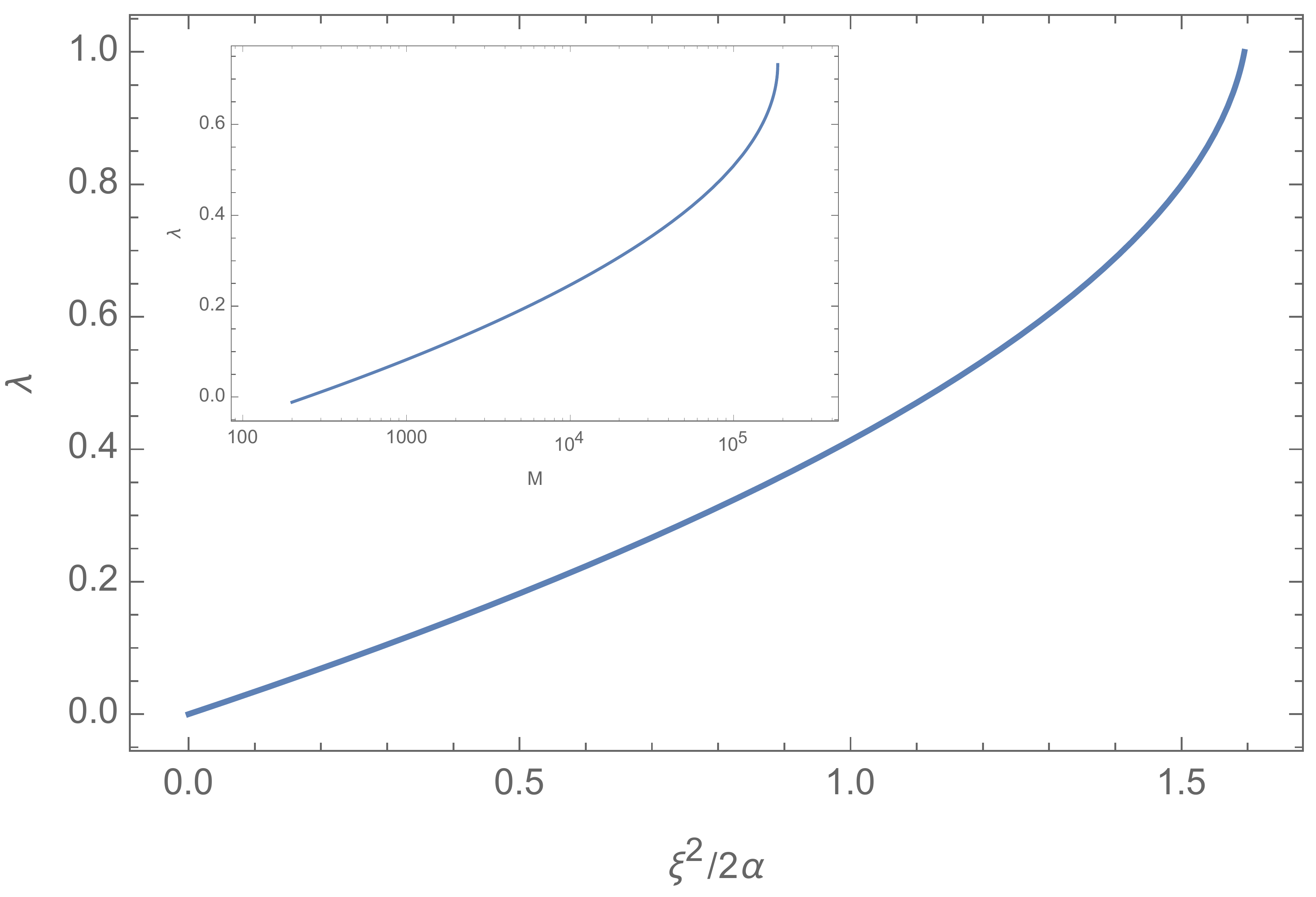}\caption{$\lambda$ versus $\xi^2/2\alpha$. The inset shows the value of the renormalization scale for a given self-coupling parameter.}\end{figure}\end{center} 

At the leading order, the Higgs effective potential \eqref{effpot} is a function of tree free parameters $\lambda, \mu,M$ and two field values $\varphi, \chi$. On other hand, there are four constrains: two extrema conditions $\partial_\varphi V_{\rm E}^{\rm eff}=0$ and $\partial_\chi V_{\rm E}^{\rm eff}=0$ and two experimental inputs $v\approx 246$ GeV and $m_h\approx 125$ GeV. We numerically solve the equations and find that there exists a family of solutions with correct values for the Higgs VEV and the Higgs pole mass. These solutions are determined by $\lambda$ and $\xi^2/2\alpha$ and presented in figure 1 (we plotted the result for $\lambda$ up to 1 above which roughly perturbative expansion breaks down). For all points on this curve we have a vacuum resembling the experimentally observed EWSB vacuum. 

It is interesting note that in this framework, the Higgs mass does not fix the Higgs self-coupling as it does in the SM. Theoretically, it can assume any value smaller or greater than the SM prediction (about 0.13). It remains to be measured in the next generation of colliders and once determined, we can make a prediction for the gravitational coupling $\xi$ (given $\alpha$ from cosmological inflationary observations). 
 
At the end, we also note that since
\be \partial_\chi\mu_{\rm eff}^2 = -\frac{\xi}{\sqrt{6}\alpha}e^{-\tilde\chi}\mpl + {\cal O}(\mu^2_{\rm eff}/\mpl),\ee
then from \eqref{chi-extremum} we analytically find that the Weyl field develops a non-zero VEV as follows
\be \chi_0 = \xi v^2\mpl^{-1} + {\cal O}(v^3\mpl^{-2}),\ee
where $v=\varphi_0$. Consequently, the effective Higgs mass parameter reads as follows
\be \mu^2 = -(\xi^2/2\alpha) v^2 + {\cal O}(v^4\mpl^{-2}). \ee 
The Higgs mass parameter indeed gets stabilized at a non-zero value which is much smaller than the Planck scale. 
 
  \begin{center}\begin{figure*}[ht]\includegraphics[scale=.5]{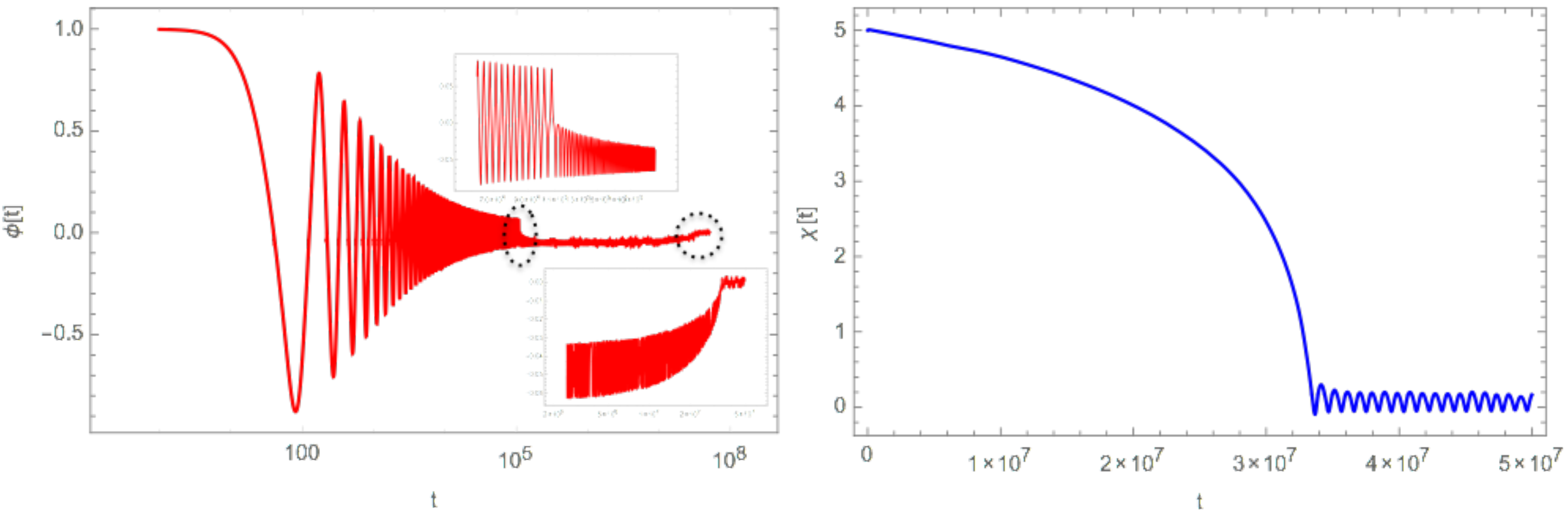}\caption{Time evolution of the Higgs (right) and the Weyl (left) fields. Field values and time are in Planck units.}\end{figure*}\end{center}

\subsubsection*{One-Loop RG Improved Effective Potential} 
In order to make the effective potential valid for large field values to check the absolute stability at high scale, we need to improve it using the renormalization group method and resum large logarithms. The one-loop RG improved effective potential for the Higgs field in the $\overline{\rm MS}$ scheme is given by \cite{Ford:1992mv,Bando:1992np,Ford:1992pn,Casas:1994qy}
\ba V_{\rm eff}[t] &=&\frac{1}{2}\mu^2(t)\varphi^2(t) +\frac{1}{4}\lambda(t)\varphi^4(t)\cr &+&\frac{1}{64\pi^2}\sum_i n_i M_i^4(t)\big[\ln\big(M_i^2(t)/M^2(t)\big)-c_i\big],\qquad\ea
which satisfies the RG equation ${\rm d}V_{\rm eff}/{\rm d}t=0$. 
The renormalization scale is given by the running parameter $t$ and some fixed scale $M$ as $M(t) = Me^t$. We choose $M(t)$ at our convenience proportional to $\varphi(t)$ where the running Higgs field is
\be \varphi(t) = e^{-\int_0^t\frac{\gamma(t')}{1+\gamma(t')}{\rm d}t'}\varphi_c.\ee
In the above, $\gamma(t)$ is the Higgs anomalous dimension and $\varphi_c$ is classical field value. The self-coupling and the mass parameter are determined through
\be \frac{{\rm d}\lambda(t)}{{\rm d}t} = \frac{\beta_\lambda(t)}{1+\gamma(t)},\quad \frac{{\rm d}\mu^2(t)}{{\rm d}t} = \frac{\mu^2(t)\beta_{\mu^2}(t)}{1+\gamma(t)},\ee
and so are the rest of the SM couplings. The one-loop improved effective potential by two-loop RG functions resums next-to-leading order logarithmic corrections \cite{Ford:1992mv,Ford:1992pn,Bando:1992np}. The RG functions can be systematically deduced from the unimproved effective potential \eqref{effpot}.

The boundary conditions are set through physical observables by demanding non-zero expectation value and experimentally measured Higgs mass as follows
\ba v(t_0) &\approx& 246e^{-\int_{M_Z}^{t_0}\gamma(t')/(1+\gamma(t')){\rm d}t'}{\rm GeV},\\
m_h^2 &=& m_h^2(t_0)e^{-2\int_{t_0}^{M_Z}\gamma(t')/(1+\gamma(t')){\rm d}t'}+{\rm Re}\big[\Pi(m_h^2)-\Pi(0)\big]\nn\\&\approx& 125 {\rm GeV},\ea
where $t_0$ is where the potential is extremized.
The RG equations for RG functions with the above boundary conditions can be numerically solved to get the improved potential. 

Again with the convention we adopted in \eqref{potential}, \eqref{eff-lambda} and \eqref{eff-mass} we follow the result of \cite{Casas:1994qy} (to where we refer the reader for details) for the improved Higgs potential. We find that absolute stability up to the Planck scale can be guaranteed if the self-coupling is greater than about $0.14$. It is indeed possible in present framework since it is not fixed by any current experimental measurements.

\subsubsection*{Higgs Dynamics in the Early Universe}
In order to explicitly follows the fields dynamics in the early Universe and how they connect to the late-time vacuum, we numerically solve the equations of motions in a homogenous background
\ba &&\ddot\chi + 3H\dot\chi + 6^{-1/2}e^{-\tilde\chi}\dot\varphi^2+\partial_\chi V^{\rm eff}_{E} = 0,\\
&&\ddot\varphi + 3H\dot\varphi - (2/3)^{1/2}\mpl^{-1} \dot\chi\dot\varphi +\partial_\varphi V^{\rm eff}_{E}  = 0,\\
&& 3 H^2 \mpl^2 = \frac{1}{2}\dot\chi^2 + \frac{1}{2}e^{-\tilde\chi} \dot\varphi^2 + V_E,\\
&&2 \dot H \mpl^2 = -\dot\chi^2 - e^{-\tilde\chi} \dot\varphi^2.\ea
In the above equations, since the fields start at large values, the improved effective potential is used. A primary study has been done in \cite{Torabian:2014nva} and here we applied to our special case and plotted the result in figure 2.

For large field values, the Higgs effective mass is big (compared to the Hubble scale) in the early times and drives it to small field values thorough dissipative coherent oscillations. When it loses its amplitude it gets trapped in one of the minima and follow oscillations about the symmetry breaking minimum. Eventually, when the energy density in the Higgs field dissipated away the Weyl field takes over. It slowly role down its potential and drives a period of cosmic inflation. Consequently through many oscillations, it relaxes to its minimum very close to zero. As we saw explicitly in the previous section, quantum fluctuations stabilized both fields at non-zero small values.

\subsection*{Conclusion}
In this paper we introduced a framework in which the weak and the Planck scale hierarchy could be explained through dimensional transmutation induced by the RG running of the couplings and cosmological evolution of the Weyl field. In this setup, the Higgs mass-squared is not a parameter to be defined by fine-tuning at the UV scale, but instead an effective dynamical variable which takes diverse values during different stages of cosmic history. It essentially  gives a manifestation of the brilliant Coleman-Weinberg idea for EWSB now assisted by gravitational couplings. 

Moreover, we found that the gravitational couplings also modifies the Higgs quartic self-coupling. As a result, the Higgs pole mass and the self-coupling are not directly related and the measurement of one them would not determine the other. Therefore, the quartic coupling can take any perturbative value including values which make the EW vacuum absolutely  stable.

\paragraph*{Acknowledgments}
MT would like to thank ICTP, Trieste for hospitality during the initial stages of this work.

\ \\$^*$ Electronic address: mahdi@physics.sharif.edu

\end{document}